\documentclass[pre,superscriptaddress,showpacs,twocolumn,endfloats*]{revtex4}
\usepackage{graphicx}

\newcommand{\Pp}{\mathrm{Prob}}
\newcommand{\osigma}{\overline{\sigma}}

\begin{document}

\title{Fluctuation theorem for constrained equilibrium systems}
\author{Thomas Gilbert}
\email[]{thomas.gilbert@ulb.ac.be}
\affiliation{Center for Nonlinear Phenomena and
  Complex Systems,
  Universit\'e Libre de Bruxelles, Code Postal 231, Campus Plaine, B-1050
  Brussels, Belgium}
\author{J. Robert Dorfman}
\email[]{rdorfman@umd.edu}
\affiliation{Department of Physics and Institute for Physical Science and
  Technology, University of Maryland, College Park, Maryland 20742}
\date{\today}

\begin{abstract}
We discuss the fluctuation properties of equilibrium chaotic systems with
constraints such as iso-kinetic and Nos\'e-Hoover
thermostats. Although the dynamics of these systems does not typically
preserve phase-space volumes, the average phase-space contraction rate
vanishes, so that the stationary states are smooth.
Nevertheless finite-time averages of the phase-space
contraction rate have non-trivial fluctuations which we show satisfy a simple
version of the Gallavotti-Cohen fluctuation theorem, complementary to the
usual fluctuation theorem for non-equilibrium stationary states, and
appropriate to constrained equilibrium states. Moreover we show
these fluctuations are distributed according to a Gaussian curve for
long-enough times. Three 
different systems are considered here, namely (i) a fluid composed of
particles interacting with Lennard-Jones potentials; (ii) a harmonic
oscillator with Nos\'e-Hoover thermostatting; (iii) a simple hyperbolic
two-dimensional map.
\end{abstract}

\maketitle

Recently some papers have appeared which discuss the applicability of
the Gallavotti-Cohen Fluctuation Theorem (FT) \cite{GC95} to near equilibrium
and equilibrium chaotic systems subject to isokinetic and Nos\'e-Hoover 
thermostats \cite{ESR05, DK05, BGGZ05}. In particular, in \cite{ESR05},
Evans \textit{et al.} consider constrained isokinetic systems of
interacting particles which are in equilibrium or near equilibrium
states and claim the fluctutations of the phase-space contraction rate of
the systems at or near equilibrium do not satisfy the FT. 

One of the issues under discussion in these papers is
whether or not the FT applies when the external forces vanish, 
which we refer to as constrained equilibrium. These systems differ from
genuine equilibrium ones because of the action of a constraint, for
instance acting so as to keep the kinetic energy constant.
For systems consisting of particles with instantaneous, elastic, hard ball
collisions the total energy is all kinetic and the constraint has no effect
in the absence of external forces, thus the resulting ensemble is the usual
microcanonical one. For other systems the kinetic and total energies
are not the same. The constraint fixes only the kinetic energy and the
total energy fluctuates. As a result, the constrained equilibrium
ensemble is not microcanonical, but has a different structure, which
we explore here.

For the sake of illustration, let us consider a fluid composed of particles
with pair-wise, central interactions, placed in a periodic box, and not
subjected to external forces. In the absence of other constraints, the
microcanonical measure is the invariant one. Now we modify the dynamics by
introducing a (time-reversible) frictional force on the particles with the
requirement that the total kinetic energy remain  constant in time, see
Eqs.~(\ref{ljq}), (\ref{ljp}) and (\ref{ljalpha}) below for a specific
example. Because of this constraint, the total energy is not conserved and
the Liouville theorem is violated. In 
such a case the phase-space flow can have non-vanishing  divergence
(point-wise), even though it vanishes in average. In other words, as long
as there is no external driving, the system relaxes to an equilibrium
state, different from the one specified by the volume measure, but nevertheless
with vanishing average phase-space contraction rate. For Anosov systems or
those satisfying the Chaotic Hypothesis, this implies the sum of the
Lyapunov exponents is identically zero. This is to say the stationary 
measure is smooth both with respect to stable and unstable
manifolds \cite{BGGZ05}. Moreover, the existence of a 
time-reversal symmetry implies that the equilibrium measure must be
symmetric under time-reversal operation, {\em i.~e.} a trajectory and its
time reverse are equally as probable. This is indeed what one expects from
an equilibrium state, whether it verifies Liouville's theorem or not. 

The aim of this paper is to provide through some elementary
considerations a characterization of the fluctuations of the phase-space
contraction rate of such constrained equilibrium systems, which we will
subsequently support through numerical studies of specific examples.
The outcome indicates that the FT does apply, in agreement with conclusions
reached by Gallavotti \textit{et al.} \cite{BGGZ05}.  

For non-equilibrium stationary states, the FT \cite{GC95} is a statement
about the asymmetric part of the probability distribution of the
phase-space contraction rate.  
Let $\osigma \equiv \langle \sigma \rangle$ denote 
the non-zero expectation value of the phase-space contraction rate of the
non-equilibrium stationary state of a given system and let $\Pp(\sigma_t =
p\osigma)$ denote the probability 
of observing, during a time interval of length $t$, a given fluctuation
of the partial average of $\sigma$ equal to $p \osigma$, where the amplitude
$p$ is a dimensionless number. The FT \cite{GC95} states that when $\osigma>0$
\begin{equation}
\lim_{t\to\infty}\frac{1}{t}\ln \frac{\Pp(\sigma_t = p\osigma)}
{\Pp(\sigma_t = -p\osigma)} = p \osigma. 
\label{GCFT}
\end{equation}
As pointed out in \cite{BGGZ05}, the derivation of this result {\em
  assumes} $\osigma > 0$ and must be reformulated if $\osigma = 0$. 

For an equilibrium system, finite-time distributions of the phase-space
contraction rate are symmetric about zero. Now let $\sigma$ denote the
point-wise divergence of a flow with the properties described above (or
equivalently 
the logarithm of the Jacobian of a time-discrete mapping with similar
properties), and let $\Pp(\sigma = a)$ denote the (equilibrium) probability
of observing a given value $a$ of $\sigma$. For an equilibrium system, the
most general FT is the following~:
\begin{equation}
\Pp(\sigma = a) = \Pp(\sigma = -a),
\label{FTeq}
\end{equation}
which holds since the equilibrium distribution is symmetric under
time-reversal (this operation changes the sign of $\sigma$). Equation
(\ref{FTeq}) is similar to detailed balance for stochastic systems.

For a system whose phase-space volumes are preserved by the evolution, the
probabilities in Eq. (\ref{FTeq}) are non-vanishing only at $a=0$, so that
no fluctuations occur. But this
is not necessarily so for constrained equilibrium systems.
If so, one can consider the ratio of the two probabilities, equal
to unity, and average $\sigma$ over some time interval without affecting
the result.
Let $\sigma_t$ denote the average of $\sigma$ over a time 
interval of length $t$ and take the logarithm of the ratio of
Eq.~(\ref{FTeq}) applied to $\sigma_t$ to obtain the expression
\begin{equation}
\lim_{t\to\infty}\frac{1}{t}\ln \frac{\Pp(\sigma_t = a)} {\Pp(\sigma_t =
  -a)} = 0, 
\label{GCFTeq}
\end{equation}
which is a corollary to Eq.~(\ref{FTeq}) and, as remarked in \cite{BGGZ05},
is the natural generalization of the Gallavotti-Cohen FT, here applied to
an equilibrium system. The statement of Eq.~(\ref{GCFTeq}) is weaker than
Eq.~(\ref{FTeq}) and not very useful for its own sake since the equilibrium
stationary state carries no asymmetric part. Nevertheless it is correct and
not in contradiction with the FT as stated in \cite{GC95}.

This observation does not support the comments by Evans {\em et al.}
\cite{ESR05} that for Anosov equilibrium dynamics the range of the
admissible fluctuations for the FT shrinks to zero. 
While it is true that the asymptotic range of admissible fluctuations
of $\sigma_t$ may shrink to zero as $t\to\infty$, 
one should still expect to measure fluctuations about the value $a = 0$ as
long as $t<\infty$. Moreover the rate at which $\Pp(\sigma_t = a)$ goes to
zero is the same as that of $\Pp(\sigma_t = -a)$, so that the ratio of the
two remains constant as $t\to\infty$. 
Of course, the right hand side of Eq.~(\ref{GCFTeq})
being zero, the result is trivial, irrespective of the asymptotic form of
the probability distribution. In other words, unlike
Eq.~(\ref{GCFT}) for non-equilibrium stationary states,
neither Eq.~(\ref{FTeq}), nor Eq.~(\ref{GCFTeq}) hold any information 
on the asymptotic form of the probability distribution of the phase-space
contraction rate; the only information provided by Eq.~(\ref{FTeq}) is that
the probability distributions are symmetric about their average value,
zero. Moreover, Eq.~(\ref{FTeq}) holds for equilibrium systems
irrespective of the details of the dynamics, whether it verifies the
Chaotic Hypothesis or not. The details of the fluctuations may 
indeed vary from one system to another, but even though, whether the
fluctuations are trivial or not, Eq.~(\ref{GCFTeq}) always
holds.   

Here we will show, through the example of an equilibrium hyperbolic
map \footnote{Unlike Anosov systems, which are assumed to be
  continuously hyperbolic, we will allow for discontinuities in the
  hyperbolicity, much like with the finite horizon periodic Lorentz Gas or
  with the baker map.} 
with non-trivial Jacobian, that the distribution of the phase-space  
contraction rate is symmetric about zero, 
with an asymptotic form given by $\Pp(\sigma_t = x) \approx
\rho_t(x)\mathrm{d}x$, with 
\begin{equation}
\rho_t(x) = \frac{t}{\sqrt{2\pi\chi^{2}}}\exp\left[
  -\frac{(xt)^{2}}{2\chi^{2}} \right],
\label{CLTanosov}
\end{equation}
where $\chi$ is a constant which depends on the specific system under
consideration.  We note that this indicates that the distribution of the
total phase space contraction accumulated in time has a variance that is
asymptotically independent of time.

We will show through two additional examples that for both Nos\'e-Hoover
thermostatted equilibrium systems (whose kinetic energy distribution is
given by the canonical ensemble distribution) and constrained, equilibrium
interacting particle systems, such as a Lennard-Jones fluid with isokinetic 
thermostat, the  asymptotic form of the phase-space contraction rate
distribution function is (i) non-trivial, (ii) consistent with the
relevant Equilibrium Fluctuation Relations, Eqs. (\ref{FTeq}-\ref{GCFTeq})
and (iii) verifies a central limit theorem similar to
Eq.~(\ref{CLTanosov}), albeit with a time-dependence in $\sqrt{t}$ instead
of $t$. 

The first example we consider is a mapping of the unit square into itself
with hyperbolic properties similar to the baker map. It is a variant of a
non-linear baker map previously introduced by us \cite{GFD99}. Namely let
$M_a$ be defined by 
\begin{eqnarray}
&&M_a(x,y) = \label{trigmap}\\
&&\quad\left\{
\begin{array}{l@{\quad}c}
\big(2\varphi_a(x),\varphi_{-a}(y/2)\big),
&0\leq x < 1/2,\\
\big(2\varphi_a(x)-1,\varphi_{-a}((y+1)/2)\big),
&1/2\leq x < 1,
\end{array}
\right.
\nonumber
\end{eqnarray}
where
\begin{eqnarray}
&&\varphi_a(x) = \label{fieldmap}\\
&&\quad
\left\{
\begin{array}{l@{\quad}c}
(1/\pi)\arctan\big[\tan(\pi x)2^{-a}\big],&0\leq x < 1/2,\\
1+(1/\pi)\arctan\big[\tan(\pi x)2^{-a}\big],&1/2\leq x < 1.
\end{array}
\right.
\nonumber
\end{eqnarray}

\begin{figure}[htb]
\centering
\includegraphics[width=8cm]{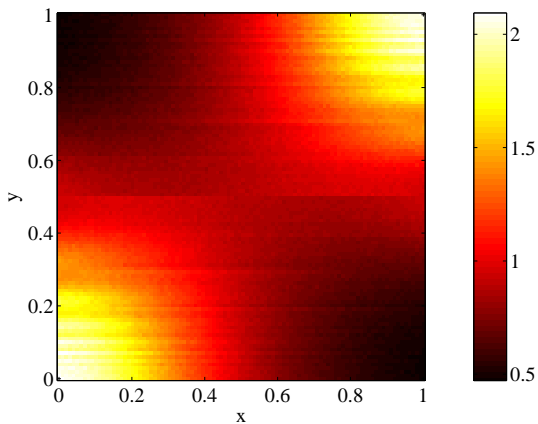}
\caption{(Color online) Numerical computation of the invariant density of
  $M_a(x,y)$ Eq.~(\ref{trigmap}) for $a=1/2$. The computation is the result
  of $10^{4}$ trajectories iterated over a time of $10^{3}$ steps. The same
  time was discarded prior to outputting trajectories in order to eliminate
  transient effects.}  
\label{fig.fieldmeas}
\end{figure}

\begin{figure}[htb]
\centering
\includegraphics[width=8cm]{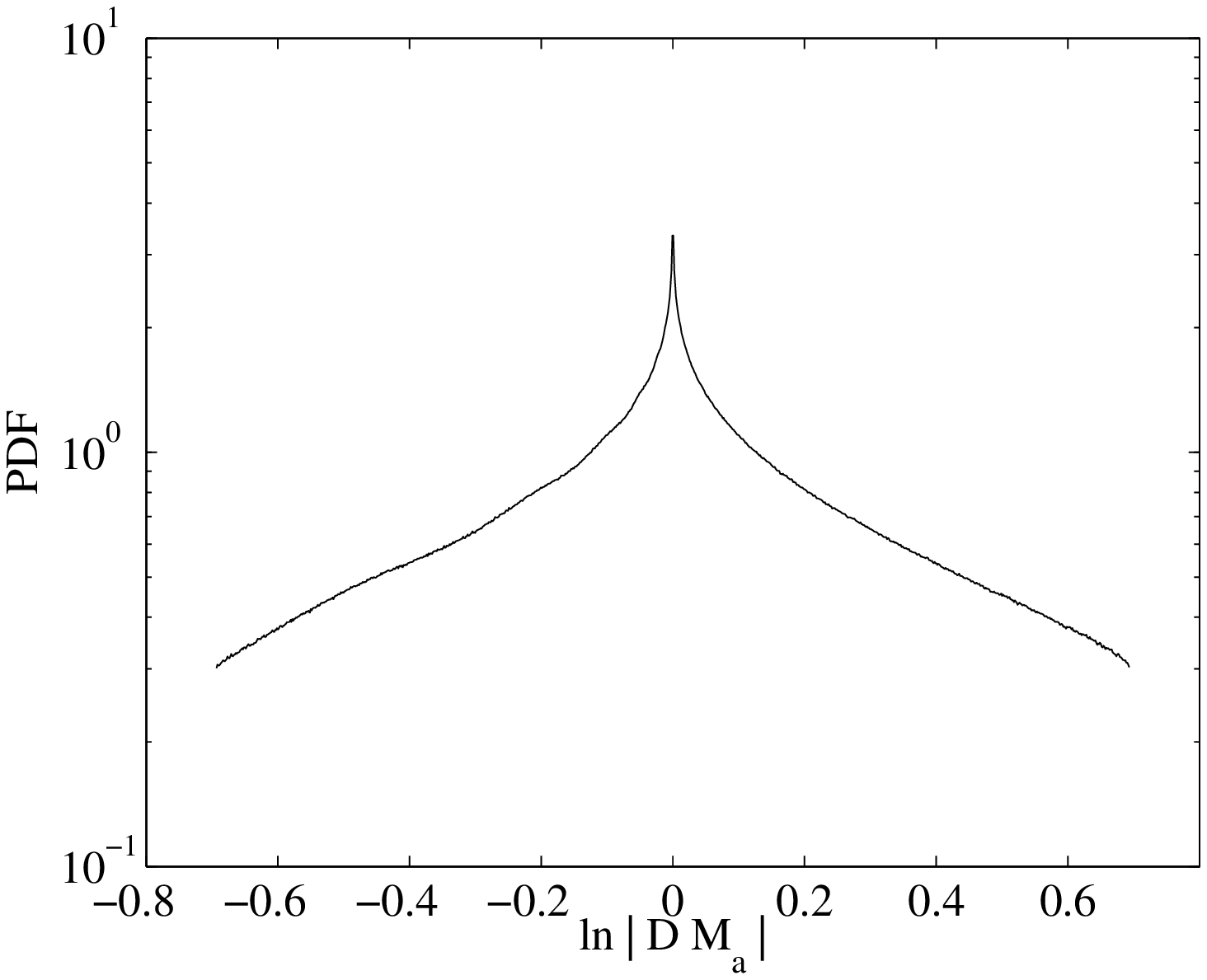}
\caption{Numerical computation of the probability density of the logarithm
  of the Jacobian of $M_a(x,y)$ Eq.~(\ref{trigmap}) for $a=1/2$. The
  computation is the result of sampling $10^{4}$ trajectories iterated over a
  time of $10^{4}$ steps. $10^{3}$ bins were used.} 
\label{fig.fieldjaco}
\end{figure}

The time-reversibility of $M_a$ under the  time-reversal operator $T(x,y) =
(1-y,1-x)$ is easy to check. Moreover, as long as $0\leq a<1$, $M_a$ is
uniformly expanding along the $x$-coordinate, while it is contracting along
the $y$-coordinate. It is smooth, {\em i.~e.} it is continuous and has
continuous derivative everywhere, but for the cut at $x=1/2$. Therefore it
shares the hyperbolic properties of the usual baker map.
The Jacobian, which we denote $|DM_a(x,y)|$, is typically
not equal to unity, but the natural invariant measure of this map turns out
to be absolutely continuous, as far as one can tell from numerical
simulations. This is illustrated in Fig.~\ref{fig.fieldmeas} where the
parameter was set to $a=1/2$. The symmetry of the invariant measure under
the time-reversal operator is clearly seen from the figure.  

\begin{figure}[htb]
\centering
\includegraphics[width=8cm]{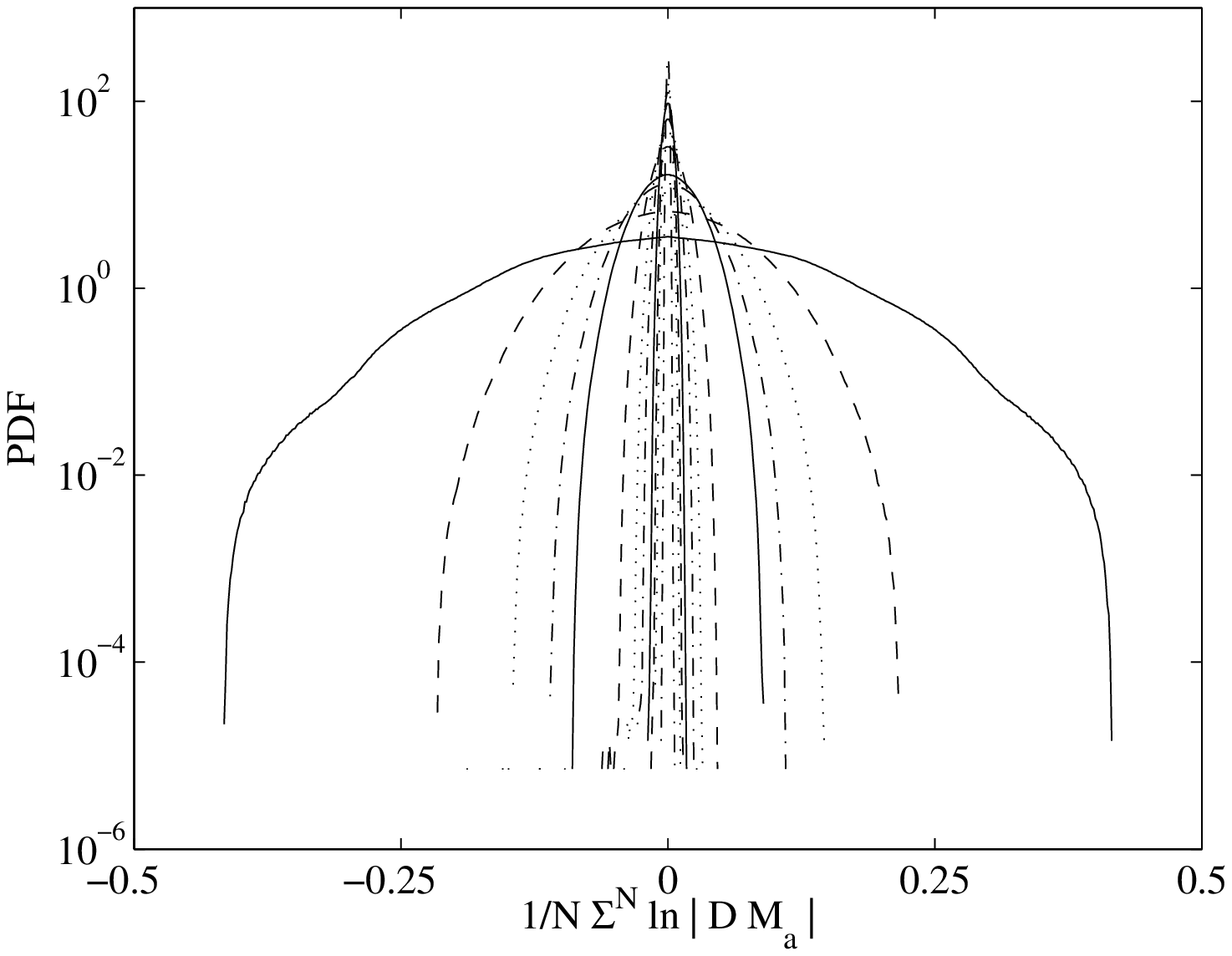}%
\caption{Numerical computation of the probability density of the
  logarithm of the time-averaged Jacobian of $M_a(x,y)$ Eq.~(\ref{trigmap})
  for $a=1/2$. The curves shown here are averaged respectively over $N=5$,
  10, 15, 20, 25, 50, 75, 100, 150, 200, 250, and 500 steps (the curves get
  narrower as the number of steps increases). The computation is the result
  of sampling $10^{4}$ trajectories iterated over a time of $10^{4}$
  steps. $10^{3}$ bins were used.}
\label{fig.fieldavg}
\end{figure}

\begin{figure}[htb]
\centering
\includegraphics[width=8cm]{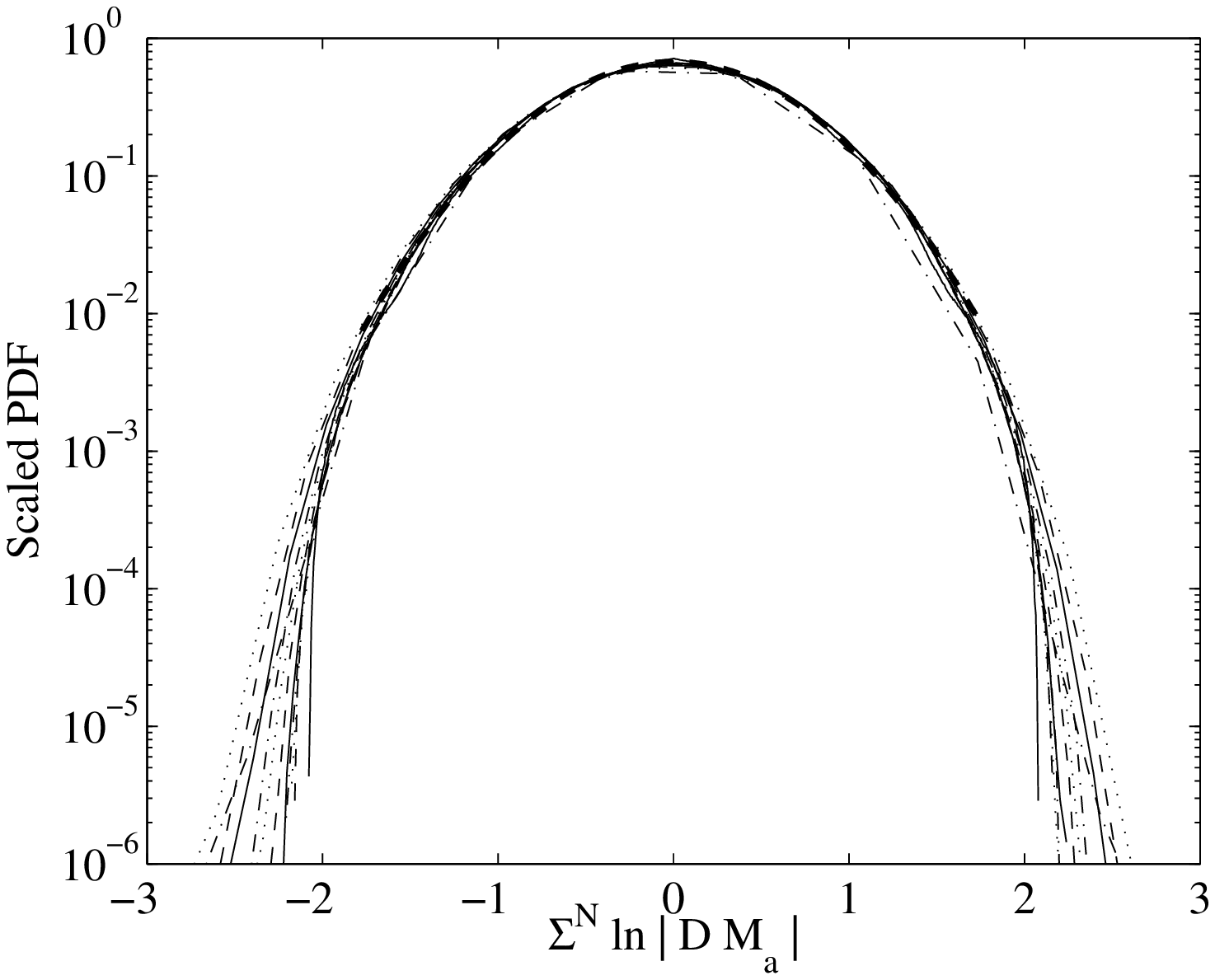}%
\caption{Same as Fig.~\ref{fig.fieldavg} with the curves rescaled according
  to Eq.~(\ref{CLTanosov}). The thick dashed curved
  is a Gaussian with standard deviation $\chi^2 = 0.35$.} 
\label{fig.fieldavgj}
\end{figure}

The smoothness of the invariant measure is consistent with the computation
of the Lyapunov exponents, which, for this value of the parameter, are 
numerically found to be $\lambda_+ = 0.6472 + o(10^{-4})$, $\lambda_- = 
-0.6472 + o(10^{-4})$, {\em i.~e.} within numerical errors
$\lambda_++\lambda_- = 0$. Figure \ref{fig.fieldjaco} shows a numerical
computation of the probability distribution function of the logarithm of
the Jacobian of $M_a$, which is symmetric about 0,  as expected from
Eq.~(\ref{FTeq}). The probability distributions of the time-averaged
Jacobian are shown in Fig.~\ref{fig.fieldavg}-\ref{fig.fieldavgj}.

The second example we consider is that of a harmonic oscillator with
a so-called Nos\'e-Hoover chain of thermostats here limited to two
thermostat variables. This generalization of the Nos\'e-Hoover thermostatting
scheme is due to \cite{MKT92}. The system is time-reversible and yields
canonical distributions of the phase variables, as shown in \cite{Ho99}.

Let $q$ denote the position variable, $p$ the momentum, $\zeta$ and $\xi$
the thermostatting variables. Taking all the oscillator parameters to be
unity, the equations of motion are (see p.~193 of \cite{Ho99})
\begin{equation}
\begin{array}{l}
\dot q = p,\\
\dot p = -q - \zeta p,\\
\dot \zeta = p^{2} - 1 - \xi\zeta,\\
\dot \xi = \zeta^{2} - 1.
\end{array}
\label{nhosc}
\end{equation}
In the absence of a thermostatting mechanism, oscillations would be
periodic of period $2\pi$, but this is no longer so if the
thermostatting mechanism is turned on.

\begin{figure}[htb]
\centering
\includegraphics[width=8cm]{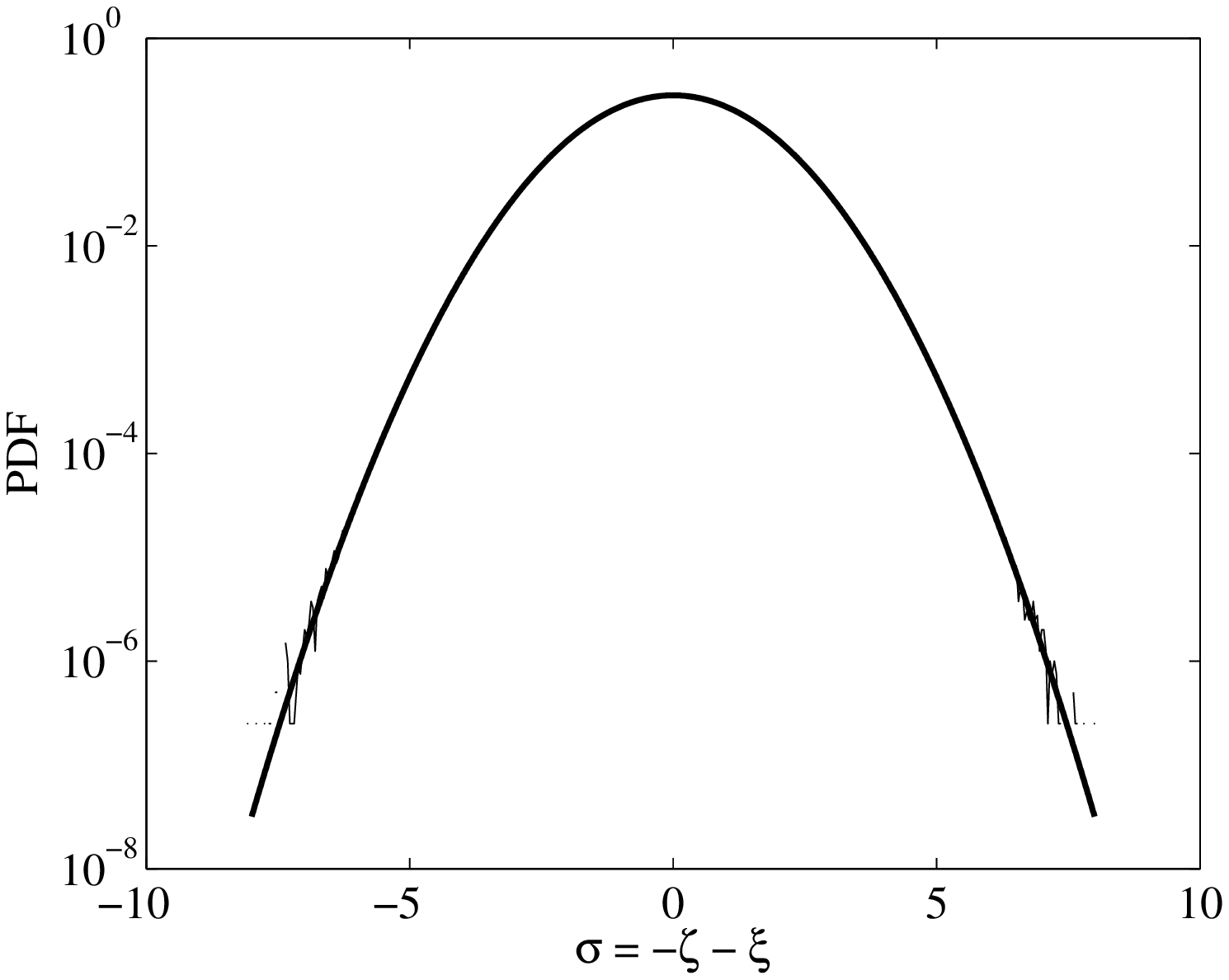}%
\caption{Numerical computation of the probability density of $-\zeta-\xi$,
  the divergence of the Nos\'e-Hoover   oscillator, Eq.~(\ref{nhosc}), for
  the trajectory with initial conditions are $q_0=0$, $p_0=1$,
  $\zeta_0=0.5$,  $\xi_0=0$. The integration is performed over a time
  $2\pi\times 10^{8}$, sampled every $2\pi$, with $500$ bins spanning
  $[-10,10]$. The thick line on top is a Gaussian with standard deviation
  $\sqrt{2}$.}  
\label{fig.nhoscdiv}
\end{figure}

\begin{figure}[htb]
\includegraphics[width=8cm]{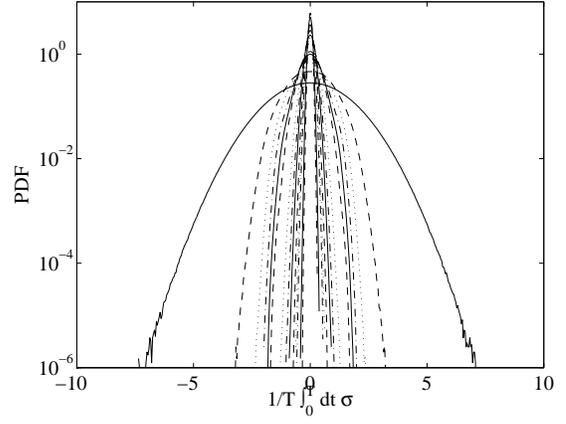}%
\caption{Numerical computation of the probability density of the
  time-averaged contraction rate $\sigma_t$, for times $T=1$, 5, 10, 15, 20,
  25, 50, 75, 100, 150, 200, 250, 500 and 750 (times $2\pi$). The
  parameters of the integration are the same as
  Fig.~\ref{fig.nhoscdiv}.}
\label{fig.nhoscdivt}
\end{figure}

\begin{figure}[htb]
\centering
\includegraphics[width=8cm]{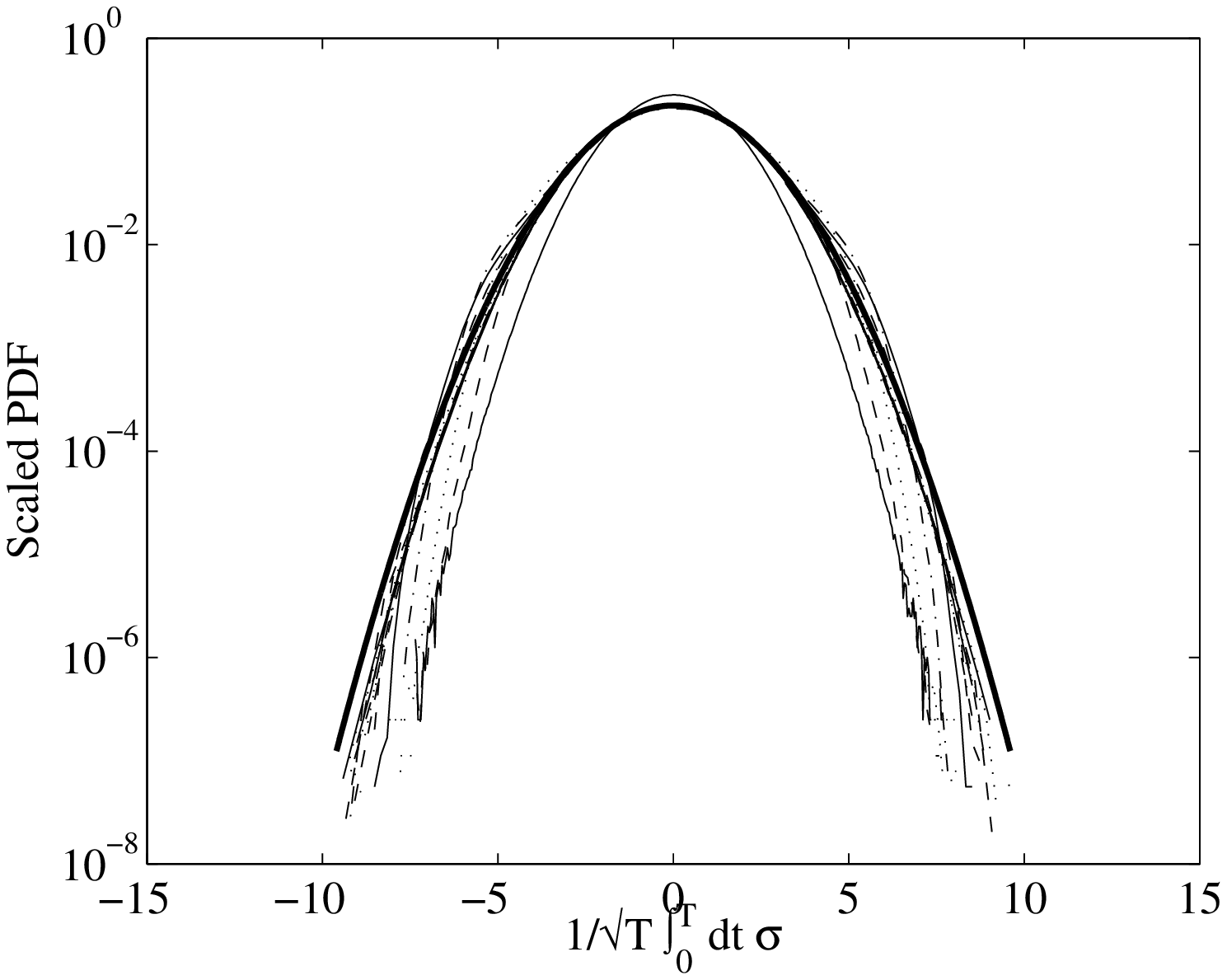}%
\caption{Same as Fig.~\ref{fig.nhoscdivt} with the curves rescaled according to
  Eq.~(\ref{CLTnhosc}). The thick curve is a Gaussian with standard
  deviation $\chi^2 = 3.2$. The agreement is excellent for times 200 and
  greater.}   
\label{fig.nhoscdivtn}
\end{figure}

The phase-space contraction rate is $\sigma = -\zeta-\xi$ whose distribution,
computed numerically with an adaptative step size 5th order Runge-Kutta
scheme \cite{NR}, is shown in Fig.~\ref{fig.nhoscdiv}. This curve is
Gaussian with standard deviation $\sqrt{2}$, consistent with the analytical
result given in \cite{Ho99}, p. 193.
Distributions of time-averaged contraction rates $\sigma_t$ are shown in
Figs.~\ref{fig.nhoscdivt}-\ref{fig.nhoscdivtn}. Deviations from an exact
Gaussian can be seen for intermediate times, which are due to surviving
correlations. However, these 
deviations disappear for long enough times, so that one retrieves a Gaussian
distribution whose asymptotic form is now given by $\Pp(\sigma_t =
x) \approx \rho_t(x)\mathrm{d}x$, with 
\begin{equation}
\rho_t(x) = \sqrt{\frac{t}{2\pi\chi^{2}}}\exp\left[
  -\frac{(x)^{2}t}{2\chi^{2}} \right].
\label{CLTnhosc}
\end{equation}
Notice the time dependence is different from Eq.~(\ref{CLTanosov}).

The third example we consider is that of a Lennard-Jones fluid with the
kinetic energy constrained to a constant value,
\begin{eqnarray}
\dot\mathbf{q}_i &=& \mathbf{p}_i,\label{ljq}\\
\dot\mathbf{p}_i &=& \sum_{j\neq i} \mathbf{F}(q_{ij}) - \alpha \mathbf{p}_i,
\label{ljp}
\end{eqnarray}
where $\mathbf{F}(q_{ij})$ denotes the force between particles $i$ and $j$
associated to the usual Lennard-Jones potential, and $q_{ij}$ is the
distance separating the particles,
\begin{eqnarray}
\mathbf{F}(r) &=& - \mathbf{\nabla} \phi(r),\\
\phi(r) &=& 4\epsilon\left[\left(\frac{r}{\sigma}\right)^{12} - 
\left(\frac{r}{\sigma}\right)^{6}\right],\\
q_{ij} &=& |\mathbf{q}_i - \mathbf{q}_j|.
\end{eqnarray}

The damping term $\alpha$ in Eq.~(\ref{ljp})
is a function of the phase-space coordinates, derived from the condition
that the kinetic energy be constant and such that the equations
(\ref{ljq}-\ref{ljp}) are time-reversal symmetric,
\begin{equation}
\alpha = \frac{\sum_{i,j\neq i} \mathbf{p}_i\cdot\mathbf{F}(q_{ij})}
{\sum_i p_i^{2}}.
\label{ljalpha}
\end{equation}
One readily sees $\alpha$ is precisely minus the divergence of the flow (up
to a multiplicative constant), and therefore the 
quantity whose probability distribution is expected to be symmetric about
zero according to Eq.~(\ref{FTeq}). This has been checked numerically for a
two-dimensional fluid of 8 particles in a box of approximately $7\times 8$
units of length squared, using the modified Verlet leapfrog algorithm, as
described in \cite{BC84}. 

\begin{figure}[h]
\centering
\includegraphics[width=8cm]{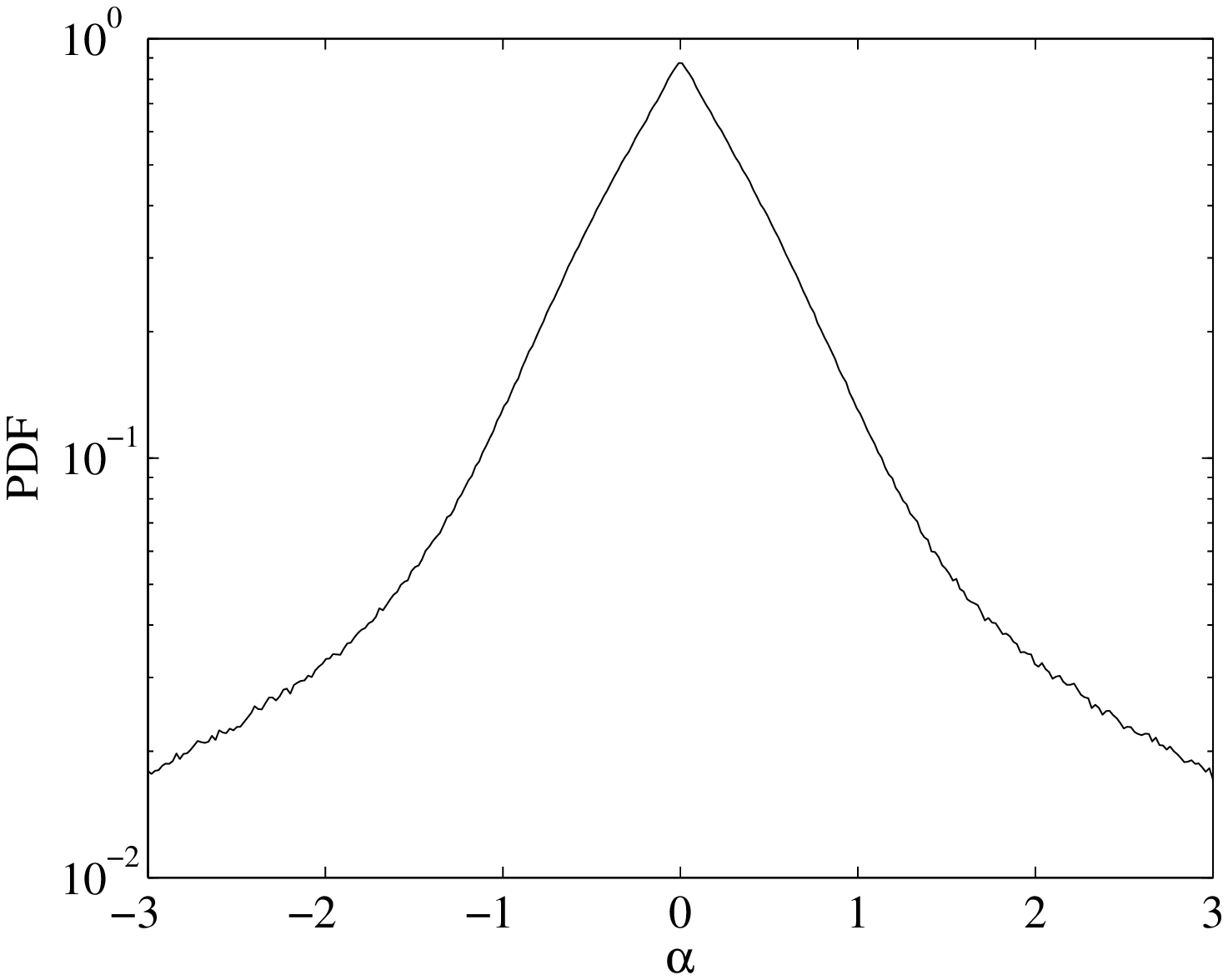}%
\caption{Numerical computation of the probability distribution of $\alpha$,
  the damping term in Eq.~(\ref{ljp}). The parameters of the simulation
  correspond to the usual choice of dimensionless parameters. $8$
  particles were used on a periodic 2-dimensional box of size approximately
  $7\times 8$.
  The temperature was set to $T=1$ and the particle masses to $m=1$. A
  number of $10^{4}$ trajectories were taken and the value of $\alpha$
  recorded for $1000$ units of time at intervals of $1$ time units. $300$
  bins were used, spanning values $\alpha \in [-3, 3]$. The average value
of $\alpha$ was computed to be $\approx 2.7 \times 10^{-4}$ with standard 
deviation $\approx 1.8 \times 10^{-3}$. About $8.7\%$ of the computed
values fell out of the range shown in the figure. }
\label{fig.ljdiv}
\end{figure}

\begin{figure}[h]
\centering
\includegraphics[width=8cm]{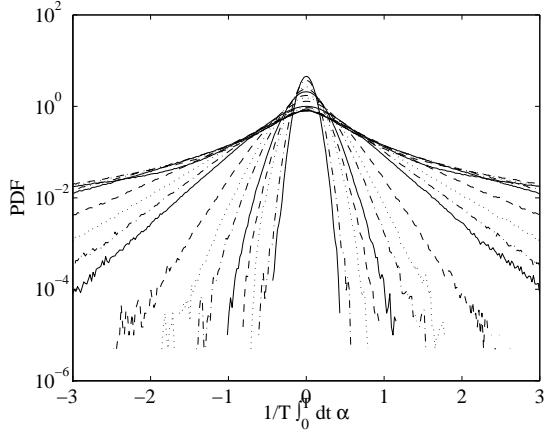}%
\caption{Same as Fig.~\ref{fig.ljdiv} for the time-average $\alpha_t =
  1/T\int_0^{T} \alpha$, for times $T = 1,$ 2, 3, 4, 5, 10, 15, 20, 25, 50,
  75, 100, 150, 200, 250, 500 and 750.}
\label{fig.ljdivt}
\end{figure}

\begin{figure}[h]
\centering
\includegraphics[width=8cm]{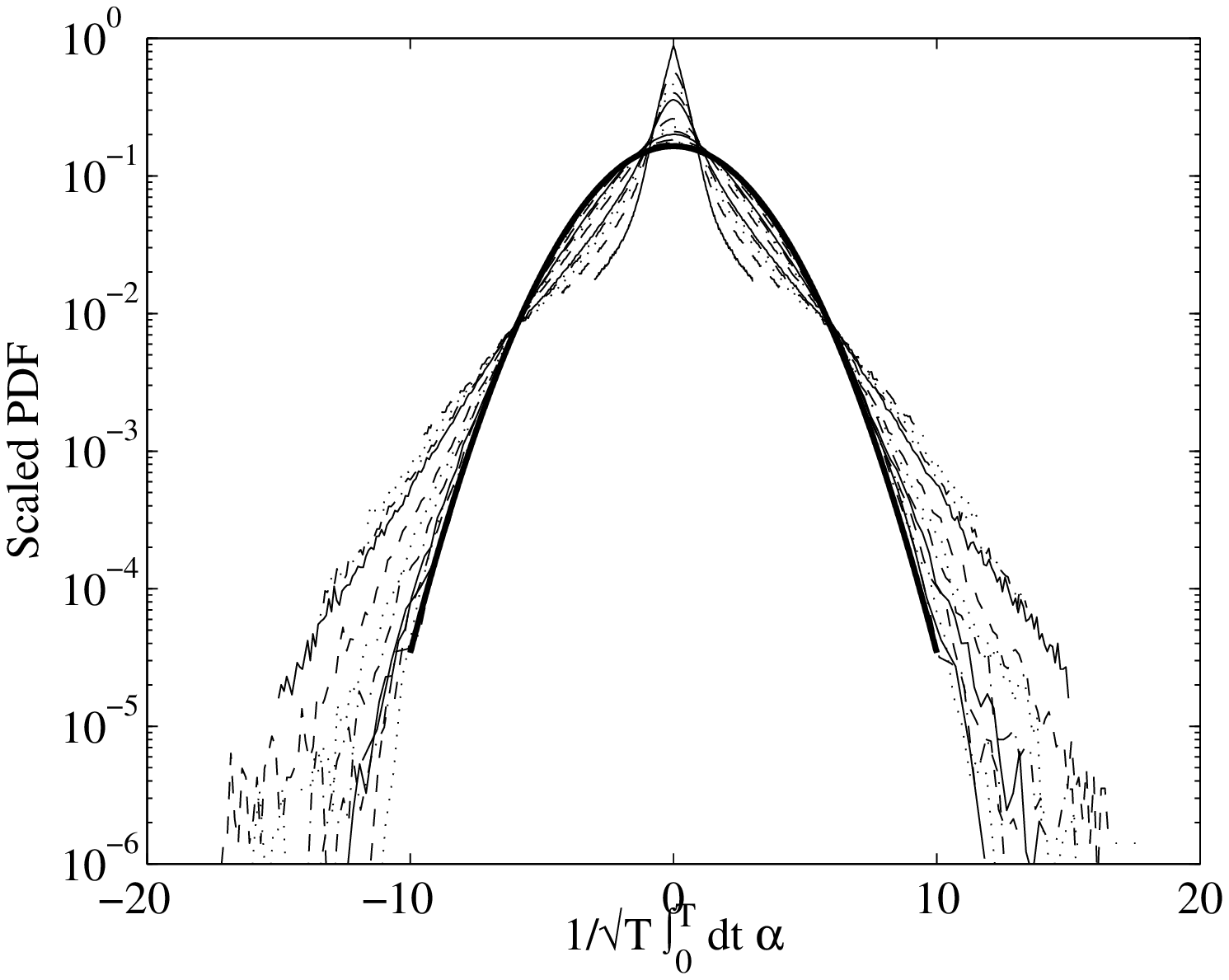}%
\caption{Same as Fig.~\ref{fig.ljdivt} with the curves rescaled
  according to Eq.~(\ref{CLTnhosc}). The thick curve is a Gaussian with
  standard deviation $\chi^2 = 5.9$. The agreement is excellent for times
  200 and greater.} 
\label{fig.ljdivtn}
\end{figure}

The resulting probability distribution of $\alpha$ is shown in
Fig.~\ref{fig.ljdiv} and the probability distributions of its time average
in  Figs.~\ref{fig.ljdivt}-\ref{fig.ljdivtn}. 

In summary, we have shown for three examples of constrained
equilibrium chaotic systems, that the probability distributions of the
phase-space contraction are symmetric about zero and verify a central limit
theorem. The time-dependence varies according to the nature of the
dynamics. It would be interesting to further investigate this property.

\begin{acknowledgments}
The authors thank G. Gallavotti and P. Gaspard, as well as D. Evans and
L. Rondoni for their comments on this work. TG is charg\'e de
recherches with the F.~N.~R.~S. (Belgium). JRD acknowledges support
from the National Science Foundation (USA) under Grant PHY 01-38697.
\end{acknowledgments}


\begin{thebibliography}{99}
\bibitem{GC95} G. Gallavotti and E.~G.~D. Cohen, Phys. Rev. Lett. {\bf 74}
  2694 (1995); J. Stat. Phys. {\bf 80} 931 (1995).
\bibitem{ESR05} D.~J. Evans, D.~J. Searles, and L. Rondoni, Phys. Rev. E
  {\bf 71} 056120 (2005).
\bibitem{DK05} M. Dolowschiak and Z. Kovacs, Phys. Rev. E {\bf 71},
  025202(R), (2005). 
\bibitem{BGGZ05} F. Bonetto, G. Gallavotti, A. Giuliani, and F. Zamponi,
  {\em Chaotic hypothesis, fluctuation theorem and singularities},
  cond-mat/0507672. 
\bibitem{GFD99} T. Gilbert, C. D. Ferguson, and J. R. Dorfman,
  Phys. Rev. E {\bf 59}, 364, (1999).
\bibitem{MKT92} G.~J. Martyna, M.~L. Klein, and M. Tuckerman,
  J. Chem. Phys. {\bf 97} 2635 (1992) 
\bibitem{Ho99} W.~G. Hoover, {\em Time reversibility, computer simulation, and
    Chaos}, (World Scientific, Singapore, 1999).
\bibitem{NR} William H. Press, Brian P. Flannery, Saul A. Teukolsky,
  William T. Vetterling, {\em Numerical Recipes in Fortran}, Second edition
  (Cambridge University Press, Cambridge, 1992).
\bibitem{BC84} D. Brown and J.~H.~R. Clarke, Molecular Phys. {\bf 51} 1243
  (1984). 
\end{thebibliography}
\end{document}